\documentclass[aps,prc,twocolumn,groupedaddress,showpacs]{revtex4}

\usepackage{graphicx}

\begin{document}
\title{Observation of Airy minimum in elastic
 and inelastic scattering of $^3$He from $^{12}$C at 50.5 and 60 MeV
 and alpha particle condensation in $^{12}$C }

\author{Sh. Hamada$^1$, Y. Hirabayashi$^2$, N. Burtebayev$^3$,
 and S. Ohkubo$^{4,5}$ }
\affiliation{$^1$Faculty of Science, Tanta University, Tanta, Egypt }

\affiliation{$^2$Information Initiative Center,
Hokkaido University, Sapporo 060-0811, Japan}
\affiliation{$^3$ Institute of Nuclear Physics, Almaty, Kazakhstan }
\affiliation{$^4$ Research Center for Nuclear Physics, Osaka University, 
Ibaraki, Osaka 567-0047, Japan }
\affiliation{$^5$University of Kochi, 5-15 Eikokuji, Kochi 780-8515, Japan  }

\date{\today}

\begin{abstract}
\par
Angular distributions for elastic and inelastic scattering of $^3$He from $^{12}$C  were 
measured at energies 50.5 and 60 MeV. The Airy minimum of the prerainbow scattering
 was clearly observed in the angular distributions for the 0$_2^+$  (7.65 MeV) state of 
$^{12}$C (Hoyle state). The experimental results were analyzed with a coupled channels
 method with double folding potentials derived from the microscopic wave functions 
for the ground 0$_1^+$, 2$^+$ (4.44 MeV), 3$^-$ (9.64 MeV) and 0$_2^+$  states. 
The analysis supports the view that the Hoyle 
state is a three alpha particle condensate with a large radius of dilute matter distribution. 
\end{abstract}


\maketitle

\section{INTRODUCTION}
The well observed $\alpha$-cluster structure of excited states of $^{12}$C has been a subject of significant
 interest \cite{Freer2007}.  Especially, much attention has been paid to the  $\alpha$-cluster structure of the 
   0$_2^+$ state at $E_x$=7.65 MeV (Hoyle state), which is 0.38 MeV above the 3$\alpha$  threshold. As early as 
1954, Hoyle showed \cite{Hoyle1954} that this level plays an extremely important role in nucleosynthesis. 
The properties of the Hoyle state in $^{12}$C determine the ratio of carbon to oxygen formed in the
 stellar helium burning process that strongly affects the future evolution of stars.
 Our complete knowledge about the unique structure of this state is far from complete \cite{Freer2008}.
 This state has a developed  3$\alpha$  cluster structure with an enlarged radius, which has been
 confirmed by theoretical cluster models such as the 3$\alpha$  generator coordinate method \cite{Uegaki1977}
 and the 3$\alpha$  resonating group method \cite{Kamimura1978}.

 The 3$\alpha$  structure of $^{12}$C was most thoroughly
 investigated by Uegaki {\it et al.} in their pioneering work \cite{Uegaki1977}, which 
showed that the 
Hoyle state has a dilute structure in a new ``$\alpha$-boson gas phase'' and clarified
 the systematic existence of
 a ``new phase''  of the 3$\alpha$ particles above the $\alpha$  threshold.  Tohsaki {\it et al.}  \cite{Tohsaki2001} suggested a condensate
 structure for the 0$_2^+$  state, where the 3$\alpha$-clusters are condensed into the lowest s-state of 
their potential. Bose-Einstein condensation (BEC) has been well established in a dilute gas
 of cold atom clusters \cite{Leggett2001}. Much effort has been devoted to understanding the alpha particle
 condensation of the Hoyle state of $^{12}$C 
 \cite{Funaki2003,Ohkubo2004,Chernykh2007,Danilov2009,Raduta2011,Freer2011B,
Zimmerman20011,Itoh20011,Manfredi2012,Freer2012,Kurokawa2004,Kanada2007,Kurokawa2007}. 
If the Hoyle state is a dilute state due to 
BEC of 3$\alpha$ particles, then it should be possible to observe physical properties characteristic
 to it such as a  huge radius. To measure such radius of the excited state is very challenging.  

The purpose of this paper is to report the measurement of the angular distributions of $^3$He 
scattering from $^{12}$C at 50.5 and 60 MeV and to observe the Airy minimum for elastic and
 inelastic scattering including the excitation to the Hoyle state.  The shift of the 
Airy minimum for the Hoyle  state toward larger angles compared with the ground state is 
found in the coupled channel analysis of the experimental data using a double folding model.
 This supports alpha particle condensation of the Hoyle state. 

The paper is organized as follows. In Sec. II the experimental procedure is given and in Sec. III
 the method used for the theoretical analysis of the experimental data is presented. 
 Section IV is devoted to the results of the experimental data, its theoretical analysis
 and discussion. The summary is   given in Sec. V.

\section{EXPERIMENTAL PROCEDURE}

The experimental measurements of angular distributions in elastic and inelastic scattering 
of  $^3$He from $^{12}$C at incident energy $E_L$= 50.5 and 60 MeV were performed at the isochronous 
cyclotron U-150M INP NNC located in Almaty-Kazakhstan, which allows accelerating protons up
 to energy 30 MeV, deuterons up to energy 25 MeV, $^3$He up to energy 60 MeV and $\alpha$-particles up
 to energy 50 MeV. Charged particles produced in the cyclotron at the source, which is located 
in the central part of the chamber in an arc discharge, by applying an appropriate gas 
(hydrogen, deuterium, helium 3 and helium 4). Their acceleration takes place in the interpolar 
space of a 1.5 meter magnet at the time of flight of particles between the dees. The energy 
spread of the beam was determined by measuring the energy spectrum of particles elastically 
scattered by a thin target of gold located in the scattering chamber. In this case, measurements 
at small angles (10$^\circ$) can be used to avoid errors due to inaccurate knowledge of the target
 thickness and angular spread of particles in the beam. For  absolute energy scale calibration, 
three alpha sources ($^{241,243}$Am+$^{244}$Cm) were used. 

  \begin{figure*}[thb]
\includegraphics[width=\linewidth]{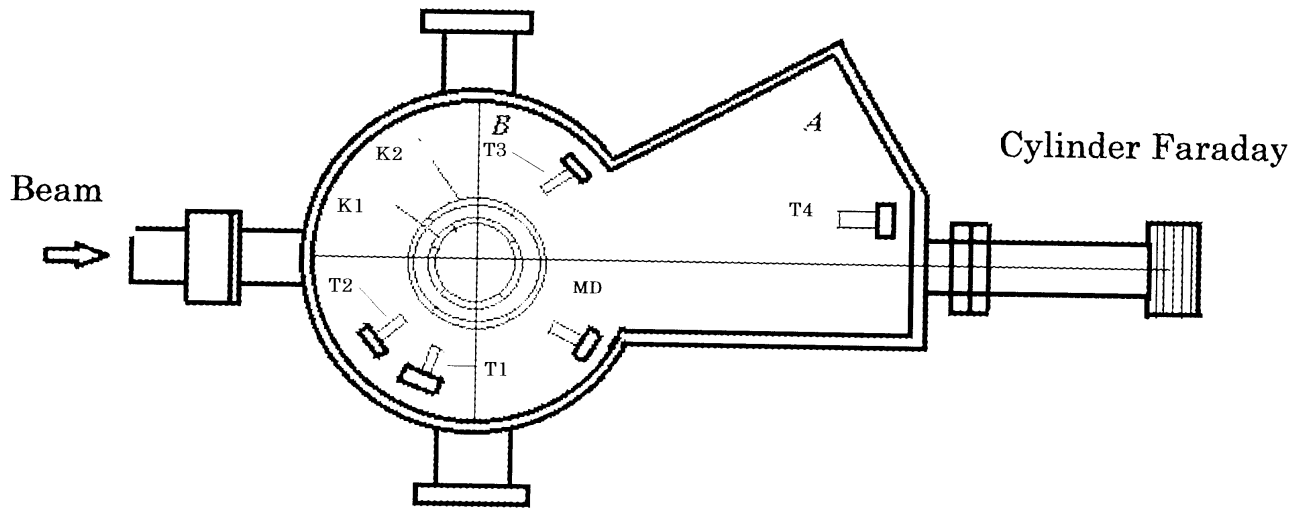}
\caption{\label{fig.1} {The layout of the scattering chamber used for the experiment.
 }
}
\end{figure*}

The $^3$He ion beam was accelerated up to energies 50.5 and 60 MeV and then directed to a $^{12}$C target
 of thickness 30$\mu$g/cm$^2$. The experiments were conducted in the scattering chamber shown 
in Fig.~1. The stainless steel scattering chamber made consists of a hollow cylinder with 
an internal diameter 715 mm, height 370 mm, and the so-called ``pocket'' A, which is an additional
 volume, elongated along the beam. In the bulk B of the chamber, there are three ($\Delta$$E$-$E$)
 telescopes of silicon semiconductor detectors, which cover the scattering angles of 10-70$^\circ$.
 A fourth telescope with independent drive, designed for covering the measurements in
 the angular range of 2-20$^\circ$, resides in  volume A.  The considerable distance from the target 
to the detectors (1000 mm) allows the load on the detecting apparatus due to the elastic
 scattering measurements at extremely small angles to be reduced by a factor of 10-15. 
A monitoring window with a diameter of 290 mm located on the top cover of the chamber 
allows visual inspection of the experimental situation (angles of the telescopes, 
the state of the target, etc.). For optimal focusing of the accelerated $^3$He ions on the
 target, two collimators of diameter 2 mm were used. The pumping system, which includes
 a high-vacuum turbo molecular pump and pre-evacuation, was tested and achieved high 
vacuum inside the chamber of about 2.3$\times$10$^{-6}$ Pa. 

The ($E$-$\Delta$$E$) method was used in the registration and identification of reaction products.
 The method is based on simultaneous measurement of specific energy losses of charged
 particles in matter $(dE/dx)$ and the total kinetic energy $E$. The method is based on the
 Bethe-Bloch theory, connecting the energy of charged particles emitted from their specific 
ionization in matter: 
\begin{equation}
\frac{dE}{dx}=\frac{kMz^2}{E}
\end{equation}
where the constant $k$  is weakly dependent on the type of particle, $M$ and $z$ are mass and 
charge of emitted particles. This relationship shows that each type of particle is 
represented by a hyperbola in 
the coordinate space ($E$,$\Delta$$E$). Thus through the simultaneous measurement of 
$E$ and $dE/dx$,
  the desired type of particle can be selected. In the telescope detectors
 ``$E$-$\Delta$$E$'', $\Delta$$E$ - detector is a surface-barrier silicon detectors firm ORTEC- thick active 
layer of 30 to 200 $\mu$m with thin inlet ($\sim$40$\mu$g/cm$^2$ Au) and outlet ($\sim$40$\mu$g/cm$^2$ Al) windows.
 The complete absorption $E$ detector is used as a stop detector.  It is manufactured by 
ORTEC and uses high-purity silicon of  thickness 2 mm.

\section{METHOD OF THEORETICAL ANALYSIS}
\begin{table}[t]
\begin{center}
\caption{The  volume integral per nucleon
pair $J_V$ , root mean square radius $<R^2>^{1/2}$, of the real 
folding potential,  and   the  parameters of the imaginary potentials in the conventional
notation in elastic and inelastic scattering of $^3$He from $^{12}$C at $E_L$= 50.5 and 60 MeV.
The normalization factor $N_R$ is fixed to 1.28.
 }
\begin{tabular}{clccrcc}
 \hline
  \hline
  $E_L$ & $J^\pi$ &$J_V$ &  $<R^2>^{1/2}$ & $W$ & $R_W$ & $a_W$  \\
  (MeV) &   &     (MeV fm$^3$) & (fm)&(MeV) & (fm) & (fm) \\
 \hline
   50.5 & $0_1^+$  &  431 &  3.58 & 6.0  & 5.5 & 0.4 \\
      & $2^+$    &  427 &  3.56 & 11.0  & 5.0 & 0.2 \\
      & $3^-$    &  489 &  3.83 & 13.0  & 5.1 & 0.2 \\
      & $0_2^+$  &  565 &  4.38 & 16.0 & 5.8 & 0.9 \\
 60 & $0_1^+$  &  420 &  3.58 & 7.0  & 5.0 & 0.4 
\\
      & $2^+$  &  415 &  3.56 & 6.0  & 5.0 & 0.2 
\\
      & $3^-$  &  476 &  3.83 & 13.0  & 5.1 & 0.2 
\\
      & $0_2^+$ &  548 &  4.38 & 12.0 & 5.5 & 0.9 
\\
 \hline
 \hline
\end{tabular}
\end{center}
\end{table}
\begin{figure*}[tbh]
\includegraphics[keepaspectratio,width=17.2cm] {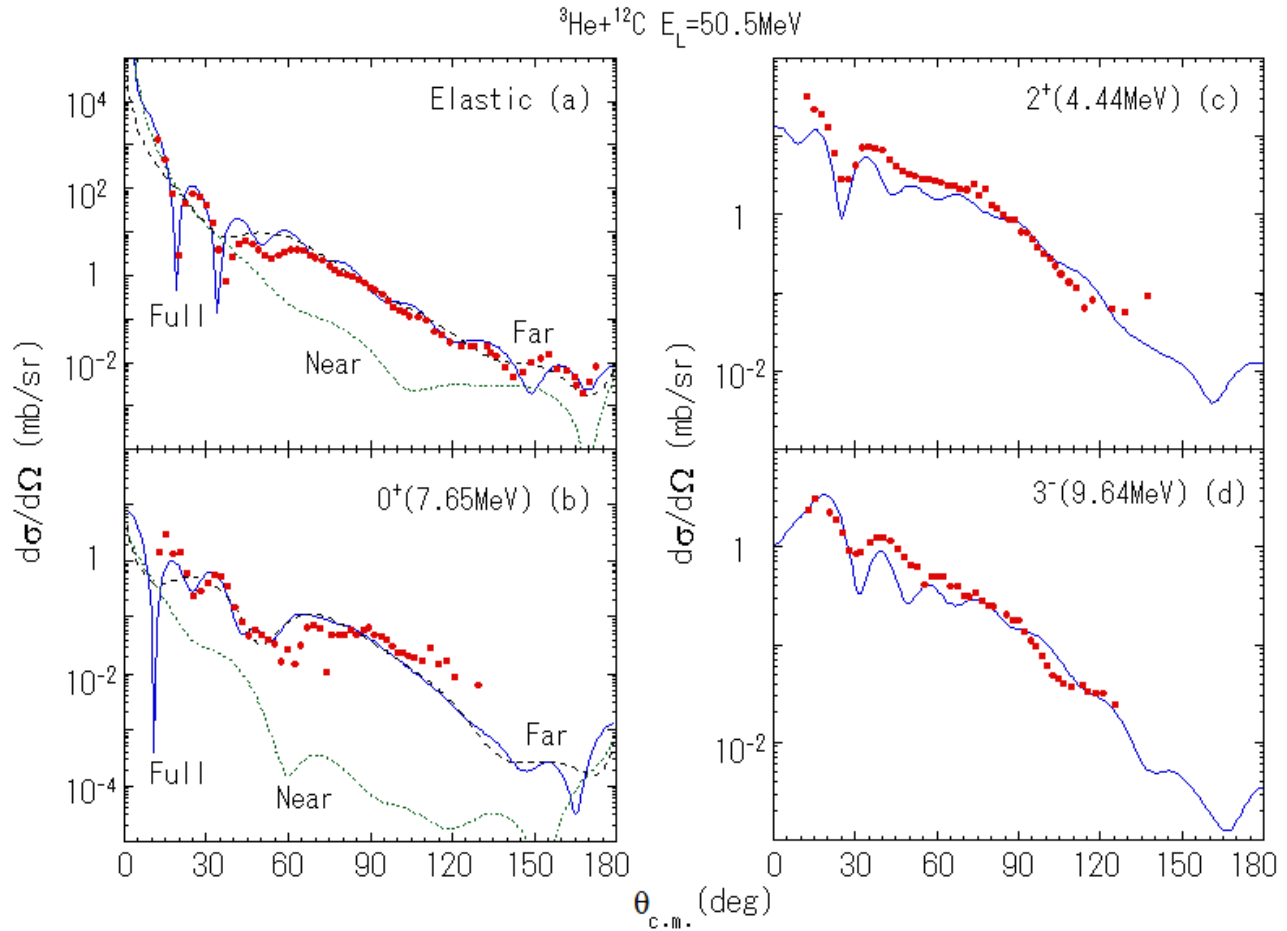}
 \protect\caption{\label{fig.2} {(Color online) 
Comparison between the experimental (points) and the calculated differential cross sections for 
elastic and inelastic scattering of $^3$He from $^{12}$C at $E_L$= 50.5 MeV: (a) ground state  (0.0 MeV), 
(b) $2^+$ (4.44 MeV), (c) $0_2^+$  (7.65 MeV) and  (d) $3^-$  (9.64 MeV) using coupled channel method.
 The calculated cross sections (solid lines) are shown 
decomposed into the farside component (dashed lines) and the nearside component (dotted lines).
 }
}
\end{figure*}

We study elastic and inelastic $^3$He+$^{12}$C scattering using the microscopic coupled
 channel method by simultaneously taking into account the 0$_1^+$ (0.0 MeV), $2^+$ (4.44 MeV),
 $0_2^+$(7.65 MeV), and $3^-$ (9.64 MeV) states of $^{12}$C. The diagonal and coupling potentials for 
the $^3$He+$^{12}$C system are calculated by the double folding model:
\begin{eqnarray}
\lefteqn{V_{ij}({\bf R}) =
\int \rho_{00}^{\rm (^3He)} ({\bf r}_{1})\;
     \rho_{ij}^{\rm (^{12}C)} ({\bf r}_{2})} \nonumber\\
&& \times v_{\rm NN} (E,\rho,{\bf r}_{1} + {\bf R} - {\bf r}_{2})\;
{\rm d}{\bf r}_{1} {\rm d}{\bf r}_{2} ,
\end{eqnarray}
\noindent
where $\rho_{00}^{\rm{ (^3He)}} ({\bf r})$ is the ground 
state density
of  $^3$He taken from Ref.\cite{Cook1981}, while $v_{\rm NN}$ denotes
 the density-dependent M3Y effective interaction (DDM3Y) \cite{Kobos1984}.
$\rho_{ij}^{\rm (^{12}C)} ({\bf r})$ represents the diagonal 
($i=j$) or transition ($i\neq j$) nucleon density of $^{12}$C
calculated in the resonating group method by 
Kamimura  \cite{Kamimura1981}.
This coupled-channel method was successfully used in the analyses of the elastic 
and inelastic scattering of $^3$He from $^{12}$C at $E_L$=34.7 and 72.0 MeV \cite{Ohkubo2007}. 
In the calculation of densities of $^{12}$C, the shell-like structure of the ground state
 0$_1^+$, $2^+$, and $3^-$  states, and the well-developed $\alpha$-cluster 
structure of the $0_2^+$  state are both well reproduced. These wave functions 
have been checked against many experimental data including charge form factors and 
electric transition probabilities involving excitation to the $0_2^+$ state [23].
 In the analysis we introduce the normalization factor $N_R$ for the real part of 
the potentials and phenomenological imaginary potentials with a Woods-Saxon form 
factor for each channel (Table I). The normalization factor $N_R$ is fixed to 1.28,
 the same as was used for a previous analysis of $^3$He+$^{12}$C \cite{Ohkubo2007}. 

\begin{figure*}[tbh]
\includegraphics[keepaspectratio,width=17.2cm] {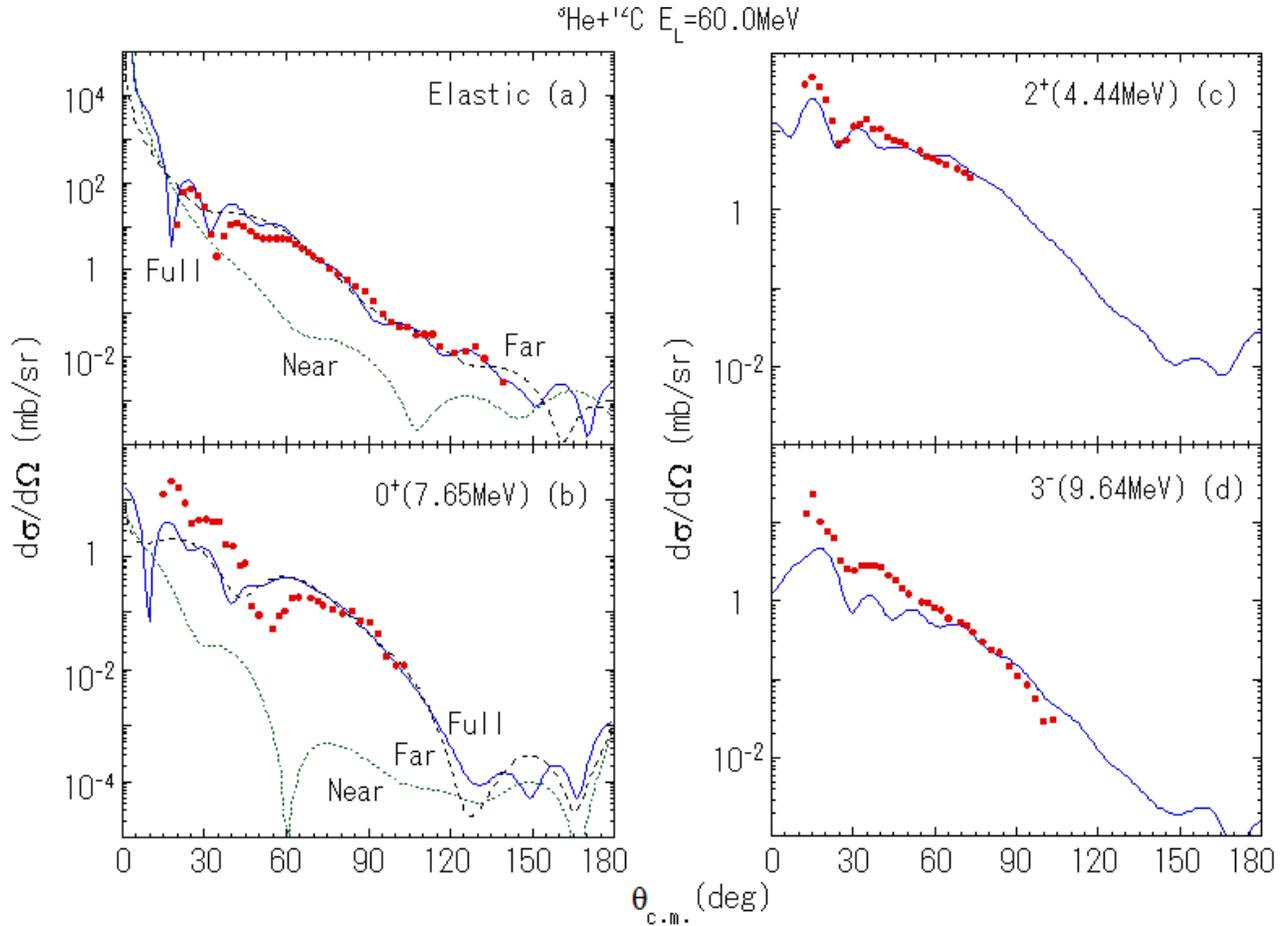}
 \protect\caption{\label{fig.3} { (Color online) Same as Fig.~2 but for $E_L$= 60 MeV.
 }
}
\end{figure*}

\begin{figure}[bht]
\includegraphics[keepaspectratio,width=8cm] {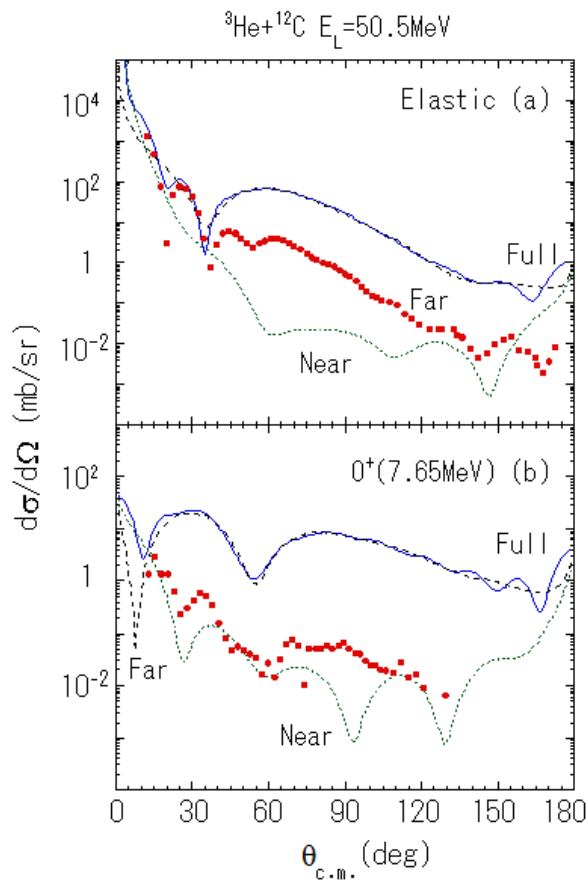}
 \protect\caption{\label{fig.4} { (Color online) 
Differential cross sections  of $^3$He scattering from $^{12}$C at $E_L$= 50.5 MeV  
calculated using coupled channel method by  switching off the imaginary potential (W=0); 
(a) elastic and (b) inelastic scattering to the  $0_2^+$  (7.65 MeV).
 The  calculated cross sections (solid lines) are shown 
decomposed into the farside component (dashed lines) and the nearside component 
(dotted lines) and compared with the experimental data (points).
 }
}
\end{figure}

\section{RESULT AND DISCUSSION}

\begin{figure}[thb]
\includegraphics[width=8.6cm]{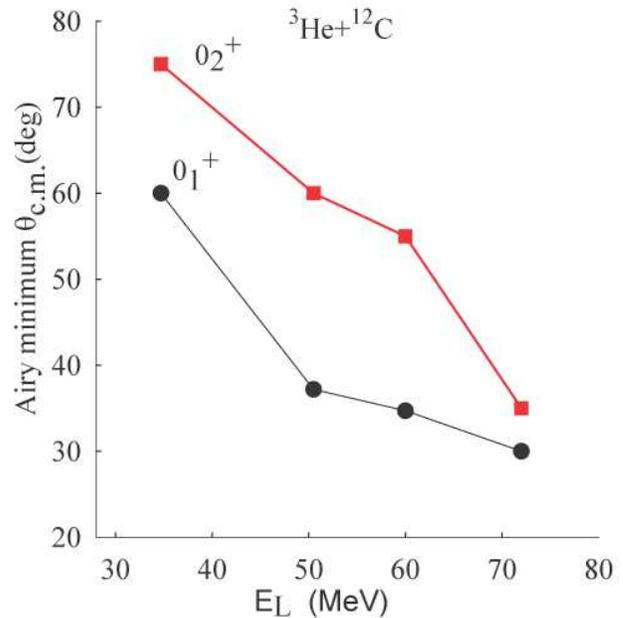}
\caption{\label{fig.5} {(Color online) Angular position of the Airy minimum $A1$ for the 
0$_1^+$ and $0_2^+$  (7.65 MeV) states of $^{12}$C in $^3$He +$^{12}$C scattering versus the 
incident energy. The data at $E_L$=34.7 MeV and 72 MeV are from Ref.\cite{Ohkubo2007}.
   The lines are only to guide the eye.
 }
}
\end{figure}

In Fig.~2, angular distributions calculated using the potentials from Table I are shown
 in comparison with the experimental data for elastic and inelastic scattering of $^3$He 
ions beam from a $^{12}$C target at energy $E_L$=50.5 MeV. The calculated cross sections are 
decomposed into the farside and nearside components. In the intermediate and large 
angular regions the scattering is dominated by the farside component, which shows 
that the scattering in this energy region is refractive. The Airy minimum can be 
 observed for elastic scattering and inelastic scattering to the Hoyle state. 
The characteristic features of the falloff of the cross sections beyond the rainbow 
angle \cite{Ohkubo2004} in the experimental angular distributions for the shell-like ground,
 $2^+$, $3^-$ states, and the $0_2^+$ state with the well-developed $\alpha$-cluster structure 
are simultaneously well reproduced. It is noted that for the ground and the
 $2^+$ states the agreement of the calculations with the data is fairly good up to
 large angles. Fig.~3 shows the comparison between the experimental data and the
 theoretical calculations  at energy $E_L$=60 MeV. The discrepancy between the experimental
 data and the calculations is seen  for the $0_2^+$ state in Fig.~2(b) and Fig.~3(b).
 One of the  causes might
  be due to the truncation of the explicit coupling to the higher excited states.
 For example, most of the imaginary potentials for the $0_2^+$ state come from the 
coupling to the 2$_2^+$ state, which has a well-developed $\alpha$-cluster structure with 
almost the same configuration as the $0_2^+$ state. 

To identify the position of the Airy minimum observed in the experimental angular distributions   
 clearly, in Fig.~4 the
theoretical angular distribution for  $E_L=$50.5 MeV calculated by switching off the 
imaginary potential  in the coupled channel method is  decomposed 
 into the farside and  nearside components. As the  scattering is caused
 by the real potential only, the features of refractive scattering such as Airy minima can 
be seen  without being obscured by absorption.
In Fig.~4(a), we see that the minimum at  35$^\circ$ in the
calculated angular distribution for  elastic scattering comes from the farside component.
Thus the  minimum at  this angle in the calculated angular distribution is assigned
 as the first order Airy minimum $A1$ caused by
refractive scattering. Although the minimum in the experimental data of elastic 
scattering is obscured by the
effect of the imaginary potential and the nearside component, we can identify uniquely 
the position of the Airy minimum in the experimental data. 
As for the inelastic scattering to the Hoyle state, in Fig.~4(b) a  clear minimum is 
seen in the
calculated angler distribution at  55$^\circ$ and the minimum is fully due to the
 farside component of the angular distribution because the contribution of the nearside component 
is negligible.  We can identify this minimum as the first order Airy minimum  
$A1$ due to refractive scattering to the Hoyle state.
The  Airy minimum $A1$ at the intermediate angular region where the farside scattering
 dominates 
 is scarcely obscured by the presence of the imaginary potential. Therefore we
 can   identify the
 $A1$ Airy minimum for inelastic scattering to the Hoyle state without ambiguity. 
 We note in Fig.~4(b) that the second order Airy minimum appears at forward angle around 
10$^\circ$ in the calculated angular distribution for inelastic scattering to the Hoyle 
state. This shows that the refraction in inelastic scattering to the Hoyle state 
is much  stronger than that in elastic scattering.   In the experimental data the 
second Airy minimum is washed out by the effect of absorption and  diffractive scattering
dominates in the very forward angular region. We have checked that the same situation appears
 for the  $E_L$=60 MeV case.

 In Fig.~2 , the first Airy minimum $A1$ at $E_L$=50.5 MeV due to  refractive 
scattering is identified  at 37$^\circ$  for
 elastic scattering and $\sim$60$^\circ$  for the $0_2^+$ state. 
In Fig.~2(b), the experimental angular distribution  to the Hoyle state shows two 
small  valleys in the intermediate angular region with a  minimum at 58$^\circ$ and
 $62^\circ$, respectively. 
It seems that these two valleys  are  fragmented from the broad valley structure of
 the Airy minimum (as seen in the theoretical  curve in Fig.4(b)) located at around
 60$^\circ$. Therefore it seems reasonable to consider that the centroid 60$^\circ$
  is a minimum due to refractive scattering, i.e., the  Airy minimum rather than
 the individual small two minima at  58$^\circ$ and $62^\circ$.
In Fig.~3 at $E_L$ =60 MeV, the first
 Airy minimum $A1$ is identified  at 35$^\circ$  for elastic scattering and 55$^\circ$ for the $0_2^+$ state.
 In Fig.~5   the position of the Airy minimum for the ground and the $0_2^+$ state 
of $^{12}$C in $^3$He+$^{12}$C scattering  at different energies (50.5, 60, 34.7 and 72 MeV)
 are summarized.  It is clearly shown that the $A1$ minimum for the $0_2^+$ state is 
shifted to a larger angle.  A more precise prerainbow Airy oscillation is seen in inelastic 
scattering to the Hoyle state than in elastic scattering. The refractive effect 
of the $0_2^+$ state is more clearly seen in the prerainbow structure at a low incident
 energy region than at a high incident energy region where a typical nuclear rainbow 
appears. This shows that the refractive effect is significantly stronger for the 
$0_2^+$ state than the ground state. 

\begin{figure}[t]
\includegraphics[width=8.6cm]{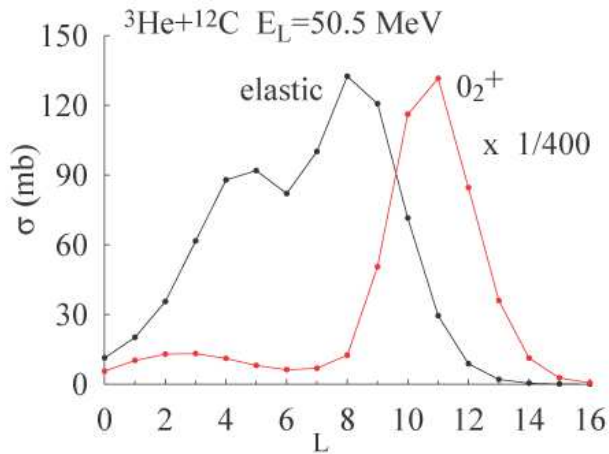}
\caption{\label{fig.6} {(Color online) Calculated partial cross sections for elastic $^3$He+$^{12}$C
 scattering and inelastic scattering to the $0_2^+$ (7.65 MeV)  state at $E_L$=50.5 MeV are shown as a function
 of the orbital angular momentum $L$ between $^3$He and $^{12}$C.
 }
}
\end{figure}

In Table I we see that the r.m.s. radius of the potential, which is a lens for 
refractive scattering, for the $0_2^+$ state is much more extended than that for the
 ground state in agreement with the dilute distribution of the density of the $0_2^+$ 
state. In Fig.~6 calculated partial cross sections scattered to the ground state 
and the $0_2^+$ state are displayed. We see that the inelastic scattering to the Hoyle
 state occurs at very large angular momenta (large radius) compared with the elastic
 scattering. These facts account for the shift of the Airy minimum of the Hoyle state
 to a larger angle and support the idea that the Hoyle state has a large radius compared
 with the normal ground state. These results are consistent with those obtained from 
refractive scattering of $\alpha$ particle scattering from $^{12}$C \cite{Ohkubo2004,Ohkubo2007,Ohkubo2007B,Ohkubo2008}.

\section{SUMMARY}

We have measured the angular distributions of elastic and inelastic scattering to
 the $2^+$ (4.44 MeV), $0_2^+$  (7.65 MeV) and $3^-$   (9.64 MeV) states of $^{12}$C by bombarding $^3$He ions on 
the thick target at energies 50.5 and 60 MeV at the isochronous cyclotron U-150M INP NNC. 
The experimental angular distributions were analyzed in the coupled channel method using
 a double folding model in which the realistic wave functions for $^{12}$C calculated in the 
microscopic cluster model were used.  It is found that the position of the Airy minimum
 in the angular distribution for the $0_2^+$ state is clearly shifted toward large angles in
 comparison with that of ground state, which indicates that the root mean square radius 
for the  state is larger than that of the ground state. This finding supports the idea 
that the $0_2^+$  (7.65 MeV) state of $^{12}$C   has a dilute density and an enlarged radius due to the 
condensation of 3$\alpha$ clusters.

\end{document}